\documentclass[%
 reprint,
superscriptaddress,
 amsmath,amssymb,
 aps,
]{revtex4-2}

\usepackage{graphicx}
\usepackage[dvipsnames]{xcolor}
\usepackage{amssymb}
\usepackage{amsmath}
\usepackage{amsbsy}
\usepackage{color}

\usepackage{bm}

\renewcommand{\selectlanguage}[1]{}

\usepackage{bm}

\usepackage{siunitx} 

\DeclareSIUnit[quantity-product = {}] \basepair{\text{bp}}
\DeclareSIUnit[quantity-product = {}] \it{\text{LAMMPS Iterations}}
\DeclareSIUnit[quantity-product = {}] \brtime{\text{$\tau_{Br}$}}

\begin{document}

\title{Fluidification of Entangled Polymers by Loop Extrusion}

\author{Filippo Conforto}
\affiliation{School of Physics and Astronomy, University of Edinburgh, Peter Guthrie Tait Road, Edinburgh, EH9 3FD, UK}

\author{Yair Gutierrez Fosado}
\affiliation{School of Physics and Astronomy, University of Edinburgh, Peter Guthrie Tait Road, Edinburgh, EH9 3FD, UK}

\author{Davide Michieletto}
\affiliation{School of Physics and Astronomy, University of Edinburgh, Peter Guthrie Tait Road, Edinburgh, EH9 3FD, UK}
\affiliation{MRC Human Genetics Unit, Institute of Genetics and Cancer, University of Edinburgh, Edinburgh EH4 2XU, UK}

\newcommand{\dmi}[1]{\textcolor{RoyalBlue}{#1}}
\newcommand{\fc}[1]{\textcolor{SeaGreen}{#1}}

\begin{abstract}
    \textbf{Loop extrusion is one of the main processes shaping chromosome organisation across the cell cycle, yet its role in regulating DNA entanglement and nucleoplasm viscoelasticity remains overlooked. We simulate entangled solutions of linear polymers under the action of generic Loop Extruding Factors (LEF) with a model that fully accounts for topological constraints and LEF-DNA uncrossability. We discover that extrusion drives the formation of bottle-brush-like structures which significantly lower the entanglement and effective viscosity of the system through an active fluidification mechanism. Interestingly, this fluidification displays an optimum at one LEF every 300-3000 basepairs. In marked contrast with entangled linear chains, the viscosity of extruded chains scales linearly with polymer length, yielding up to 1000-fold fluidification in our system. Our results illuminate how \textit{intrachain} loop extrusion contributes to actively modulate genome entanglement and viscoelasticity \emph{in vivo}. }
\end{abstract}

\maketitle


\section{Introduction}

How chromosomes are packaged within the cell while remaining accessible to transcription, replication and segregation remains one of the most fascinating unsolved problems in physics and biology. Chromosome conformation capture (3C) and related techniques~\cite{Lieberman-Aiden2009,Beagrie2017} have revealed that chromosomes are folded into so-called territories, compartments and ``Topologically Associated Domains'' (TADs)~\cite{Cremer2001,brackley2019predictive,Dekker2013,Nuebler2018}. Among the most important processes dictating chromosome folding in both interphase and mitosis is loop extrusion, performed by so-called Loop Extruding Factors (LEFs), such as cohesin, condensin and SMC5/6~\cite{nasmythCohesinCatenaseSeparate2011,Alipour2012,Fudenberg2016,Sanborn2015a,davidsonDNALoopExtrusion2019,Ganji2018,Pradhan2022,camaraSimpleModelExplains2021,Vian2018,Gibcus2018,Conte2022,Pradhan2022}.
Most of the current experimental techniques either study static snapshots of LEF-mediated chromosome conformation \emph{in vivo}~\cite{Gibcus2018}, or dynamic LEF-mediated looping process on tethered single DNA molecules \emph{in vitro}~\cite{Ganji2018}; and only very recently it was possible to track the behaviour of individual chromosome loci under the effect of loop extrusion~\cite{gabrieleDynamicsCTCFCohesinmediated2022}. Due to this, we still lack a quantitative understanding of how LEFs modulate chromosome dynamics and entanglements in the dense, crowded and entangled environment of the cell nucleus. 
To tackle this question we perform large-scale Molecular Dynamics simulations of entangled fluids of linear polymers under the action of LEFs. Specifically, we study how loop extrusion affects polymer conformation, dynamics and viscoelasticity during mitosis-like and interphase-like stages of loop extrusion, by modulating their number, processivity and turnover.

First, we find that (exclusively intrachain) loop extrusion induces a transition from linear polymers to bottle-brush-like structures, characterised by large grafting density and side-chain length controlled by the number and processivity of the LEFs, in line with previous works in dilute conditions~\cite{goloborodkoChromosomeCompactionActive2016,Fudenberg2016,baniganLimitsChromosomeCompaction2019,polovnikov_crumpled_2023}. The formation of such structures reduces the entanglement between chains due to steric interactions between the loops and entropy maximisation -- an effect dubbed ``entropic repulsion''~\cite{markoPolymerModelsMeiotic1997,Paturej2016}, in turn leading to the fluidification of the system. Second, we discover that whilst as little as $\sim$ 2 - 100 LEFs per 30 kbp of chromatin is enough to induce a significant reduction in entanglement, a larger number of LEF is not as effective due to a the reduced repulsion of side-chains and stiffening of the chain backbone. We find that extrusion enables an ``active fluidification'' process that can reduce the viscosity of long extruded polymers by three orders of magnitude with respect their non-extruded equivalent. Finally, we show that even LEFs with binding/unbinding kinetics can drive active fluidification in entangled fluids. Our work is different from other simulations on LEF-mediated disentanglement of polymers~\cite{goloborodkoChromosomeCompactionActive2016,Racko2018,Orlandini2019,polovnikov_crumpled_2023,Chan2024pnas,paturej_cyclic-polymer_2023} as we focus on the bulk viscoelastic behaviour of dense polymer solutions, rather than structure of loop-extruded polymers in dilute conditions. Our work not only numerically confirms recent theoretical arguments suggesting that loop extrusion yields an ``entanglements dilution''~\cite{polovnikov_crumpled_2023,Chan2024pnas}, but it also quantifies the degree of disentanglement in dense solutions through primitive path analysis and, more importantly, its impact on the rheology of the system.
More specifically, our results suggest that by varying the number and processivity of LEFs, the cell may be able to finely regulate entanglements between chromosomes and, in turn, the nucleoplasm effective viscoelasticity, which could be tested in large-scale imaging and spectroscopy experiments~\cite{Zidovska2013,Saintillan2018,Caragine2018}.

\section{Methods}

\subsection{Simulation details}

We model entangled DNA as semiflexible Kremer-Grest linear polymers~\cite{Kremer1990} with $N=250$, $N=500$, $N=1000$ and $N=1500$ (unless otherwise stated) beads of size $\sigma$. The beads interact with each other via a truncated and shifted Lennard-Jones potential,
\begin{equation}\label{eq:LJ}
    U_{\rm LJ}(r) = \left\{
    \begin{array}{lr}
        4 \epsilon \left[ \left(\dfrac{\sigma}{r}\right)^{12} - \left(\dfrac{\sigma}{r}\right)^6 + \dfrac14 \right] & \, r \le r_c \\
        0                                                                                                           & \, r > r_c
    \end{array} \right. \, ,
\end{equation}
where $r$ denotes the separation between the beads and the cut-off $r_c=2^{1/6}\sigma$ is chosen so that only the repulsive part of the potential is used. Nearest-neighbour monomers along the contour of the chains are connected by finitely extensible nonlinear elastic (FENE) springs as,
\begin{equation}
\scriptsize
\label{eq:Ufene}
        U_{\rm FENE + LJ}(r) = \left\{
        \begin{array}{lcl}
                -0.5kR_0^2 \ln\left(1-(\frac{r}{R_0})^2\right) + U_{\rm LJ} & \ r\le R_0 \\ \infty & \
                r> R_0                                                &
        \end{array} \right. \, ,
\end{equation}
where $k = 30\epsilon/\sigma^2$ is the spring constant and $R_{0}=1.5\sigma$ is the maximum extension of the elastic FENE bond. This choice of potentials and parameters is essential to preclude thermally-driven strand crossings and therefore ensures that the global topology is preserved at all times~\cite{Kremer1990,Tubiana2024}. Finally, we add bending rigidity via a Kratky-Porod potential, $U_{\rm bend}(\theta) = k_\theta \left(1 - \cos \theta \strut\right)$,  where $\theta$ is the angle formed between consecutive bonds and $k_\theta=5 k_BT$ is the bending constant, thus yielding a persistence length $l_p=5 \sigma$, corresponding to 50 nm in our grained model. 
We chose these parameters to facilitate the comparison with \textit{in vitro} experiments, i.e. to model the behaviour of naked DNA at physiological salt condition resulting in a screening length of 10 nm and a persistence length of 50 nm.  Each bead's motion is then evolved via the Langevin equation
\begin{equation}
    m \dfrac{d v_i}{d t} = - \gamma v_i  - \nabla U + \sqrt{2 k_BT \gamma_i} \eta
\end{equation}
along each Cartesian component. Here, $\gamma$ is the friction coefficient, $m$ the mass of the bead, $U$ the sum of the potentials acting on bead $i$ and $\sqrt{2k_BT \gamma} \eta$ a noise term that obeys the fluctuation-dissipation theorem, thus respecting the formula \begin{equation}
\langle \eta_{i}^{\alpha}(t) \eta_{j}^{\beta}(s) \rangle = \delta(t-s)\delta_{ij}\delta_{\alpha \beta}
\end{equation}
along each Cartesian component (Greek letters). The numerical evolution of the Langevin equation is done with a velocity-Verlet scheme with $dt = 0.01 \tau_{LJ}$ with $\tau_{LJ} = \tau_{Br} = \sigma \sqrt{m/\epsilon}$ in LAMMPS~\cite{Plimpton1995}.

Four different systems were considered during this work: the first one consists in a solution of 50 linear polymers having size 1000 in a cubic box of side length 80.6 (Fig.~\ref{fig:fig1}g), achieving a monomer density of  $\sim$10\%, equivalent to a volume fraction of $\phi=0.05$. We maintained the same monomer density on systems containing polymers long  250$\sigma$, 500$\sigma$, 1500$\sigma$ by reshaping the box size respectively to 51$\sigma$, 64$\sigma$ and 92$\sigma$. 

\subsection*{Modelling loop extrusion}

\begin{figure}
    \includegraphics[width=\linewidth]{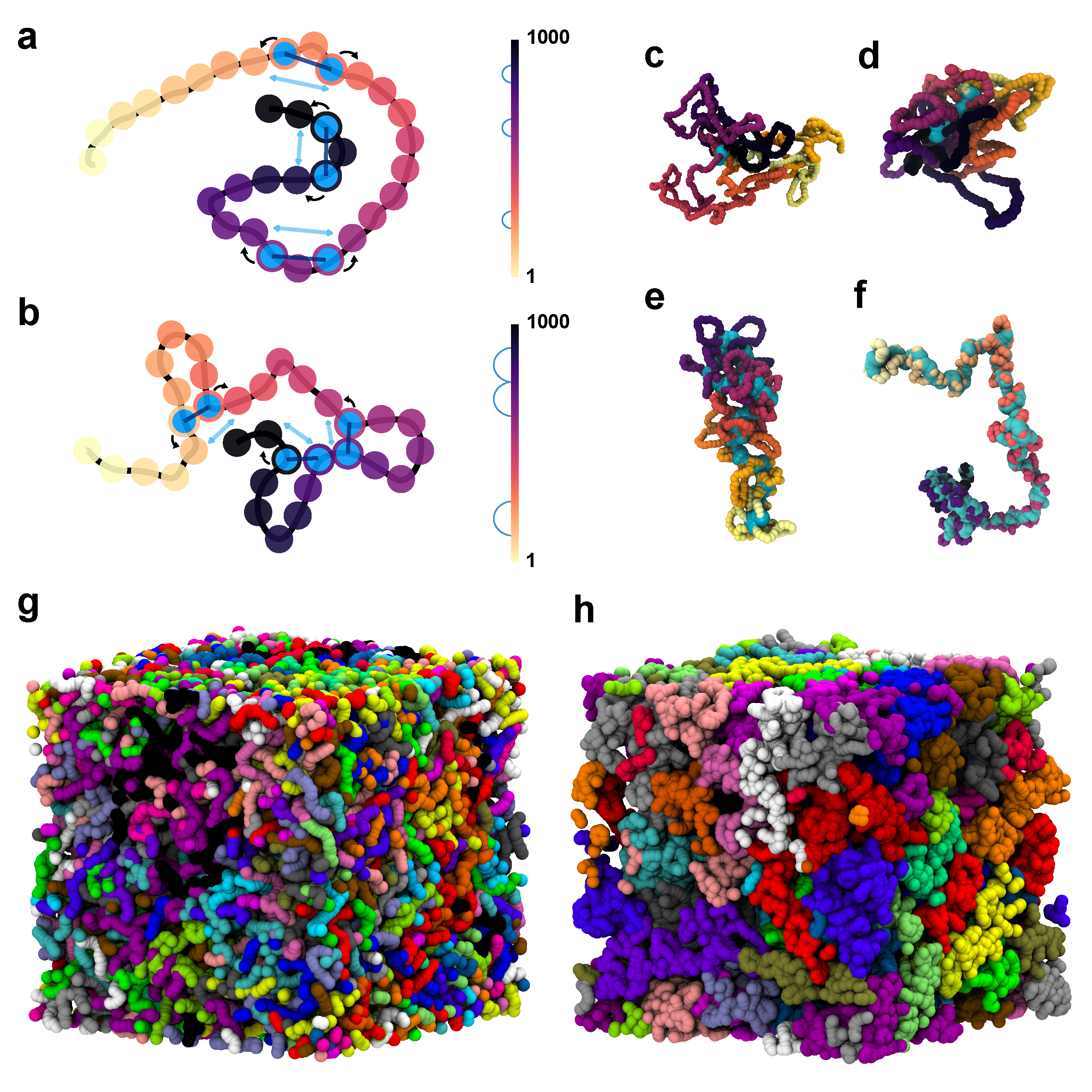}
    \caption{ \textbf{a-b} Sketch of our LEF algorithm on a coarse-grained polymer. Each LEF stops when it meets another LEF along the chain. \textbf{c-f} Snapshots of fully-extruded polymers with $n_{LEF}$ = (e) 2, (f) 10 (g) 100 and (h) 200. Cyan beads represent LEFs. \textbf{g} Snapshot of the system, consisting of $M=50$ linear polymers of length $N=1000$ in a box of size $L=80.6 \sigma$ and periodic boundary conditions. \textbf{h} Snapshot of the system after intrachain extrusion with an average of 100 LEFs per polymer. In panels \textbf{a-f} the color gradient represents the bead index, while in panels \textbf{g-h} the colors represent different polymers.}
    \label{fig:fig1}
\end{figure}

Our loop extrusion model was inspired by previous works~\cite{Fudenberg2016,goloborodkoChromosomeCompactionActive2016}. In these models, loops are formed by temporary bonds joining two beads, thus generating closed rings emerging from the polymer backbone. Loop extrusion is then achieved by shifting each bond to the adjacent beads on both sides of the bond, as shown in Fig.~\ref{fig:fig1}a. Bond-shifting introduces energy into the system, mirroring the ATP hydrolysis cycle in SMC proteins and it breaks detailed balance. However, most models in the literature allow non-physical extrusion as the bond shift is performed irrespectively of the distance between the newly selected beads. This is often possible thanks to the use of unbounded harmonic bonds, allowing large distances between the loop ends and possibly leading third segments to pass through the bonded segments. We argue that this non-physical feature should be avoided as we expect SMC complexes to block possible strand passages in between their ends, and that extrusion should take into account the geometry and topology of the DNA molecule~\cite{Orlandini2019}.
Therefore, we developed a customized version of a LAMMPS ``fix'' module publicly available at \url{https://git.ecdf.ed.ac.uk/taplab/smc-lammps}, and used the version v30\_06\_23 for this work. Specifically, we attempt extrusion steps with a fixed frequency $f_{att}$, chosen at the beginning of the simulation. On top of that, we define a success probability $f_{prob}$ for the extrusion step, that adds up to the geometry check. Then, the distance between new LEFs' ends is computed, and the step is accepted only if its value is smaller than a fixed cut-off $r < 1.2 \sigma$.
Each LEF attempts an effective step with frequency $f_{eff} = f_{att}f_{prob}$, in turn slowing down the actual extrusion speed along the polymer because of conformational entropy.
All the simulations in this paper are performed with $f_{eff} = \qty{1e-3}{\per\brtime}$.
If during the extrusion process two LEFs meet on one end, extrusion proceeds only on one side, as displayed in Fig. ~\ref{fig:fig1}b. Consequently, extrusion runs for each LEF until one of its ends neighbours another LEF's end or reaches the polymer ends. 

In practice, we implement loop extrusion by initialising a given number of LEFs by choosing random triplets of beads belonging to the polymers in solution. This provides each polymer with a total number of bound LEFs on average equal to $n_{LEF}$. However, to neglect the presence of unextruded polymers we deploy at least one LEF on every polymer.
To prevent integration errors related to the bond length, we initially model LEFs with a harmonic bond, with potential
\begin{equation}\label{eq:harmonic}
    U_{\rm harm}(r) = A(r-R_0)^2 \, ,
\end{equation}
where $A=100$ and $R_0=1.1\sigma$. 
After the first step, such bond is replaced by a FENE bond with $k=10$ and $R_0=1.7\sigma$ (see Eq. 2). We choose a larger maximum extension for the elastic FENE bond and a softer spring constant to avoid bond breaking caused by sudden movement of the bonds during extrusion.
In contrast with recent works~\cite{bonatoThreedimensionalLoopExtrusion2021,Ryu2021}, here we consider purely intrachain loop extrusion with no bridging interaction between LEFs (Fig.~\ref{fig:fig1}a-b). 

\subsection{Mean Squared Displacement and Radius of Gyration}
The mean squared displacement(MSD) at time $\tau$ measures the motion of a polymer segment with respect to the initial position over a time $\tau$. It is computed as \begin{equation}
{\rm MSD(\tau)} =  \frac{1}{N} \frac{1}{T-\tau}\sum_{t=0}^{T-\tau} \sum_{i=1}^{N} \left[ x^{i}(t+\tau)-x^{i}(t) \right]^2
\end{equation}
For entangled linear polymers we expect the MSD of the polymer's center of mass (CoM) ${\rm MSD(\tau)} \sim \tau^{\frac{1}{2}}$ for short timescales and ${\rm MSD(\tau)} \sim \tau$ on long timescales~\cite{Halverson2011dynamics}. The squared radius of gyration of a polymer is computed as

\begin{equation}
R_g^2 = \frac{1}{N} \sum_{k=1}^{N} \left[ r_k - r_{mean} \right]^2
\end{equation}
where $r_{mean}$ defines the center of mass of the polymer.

\subsection{Primitive Path Analysis}

Primitive path analysis (PPA) is a method to compute the entanglement length in a solution of polymers~\cite{Everaers2004}. We apply PPA to 10 different restart configurations for each simulated system, sampled at times separated by $\qty{2e5}{\brtime}$ to obtain uncorrelated states. The protocol used for PPA involves disabling intrachain interactions while preserving interchain interactions and keeping the polymers' ends fixed in space. To measure the entanglement between the backbones we remove the beads belonging to the loops, whereas in regular PPA the bonds forming loops are deleted. The temperature of the Langevin thermostat is set at $T = 0.001 \epsilon/k_B$ and the simulation is run for $5 \cdot 10^5$ time steps. We perform 5 of these simulations for each sampled configuration. The systems thus obtained display a collapse of the polymers over the key entanglement points, from which the entanglement length can be computed as \begin{equation}
L_e = \frac{D_{e2e}^2}{N_{beads} \cdot {\overline{D_{bond}}^2}}
\end{equation}
where $D_{e2e}$ represents the end to end distance of the polymer, and $\overline{D_{bond}}$ the average bond length in the polymer.


\subsection{Green-Kubo calculation}

The stress-relaxation modulus G(t) is calculated as
\begin{equation}
G(t) = \frac{V}{3k_bT}\sum_{\alpha \neq \beta} \bar{P}_{\alpha \beta} (0) \bar{P}_{\alpha \beta} (t)\, ,\end{equation}

where ($\bar{P}_{\alpha \beta} = \bar{P}_{xy} \, \mathrm{and} \, \bar{P}_{xz} \, \mathrm{and} \, \bar{P}_{yz}$) represents the off-diagonal components of the stress tensor. Specifically, we get those components as
\begin{equation}
\bar{P}_{\alpha \beta}(t) = \frac{1}{t_{avg}}\sum_{\Delta t = -\frac{t_{avg}}{2}+1}^{t_{avg}}P_{\alpha \beta}(t+\Delta t)\, , \end{equation}
\begin{equation}
P_{\alpha \beta}(t) = \frac{1}{V}\left(\sum_{k=1}^{NM} m_k v_k^\alpha v_k^\beta +\frac{1}{2}\sum_{k=1}^{NM}\sum_{l=1}^{NM} F_{kl}^\alpha r_{kl}^\beta \right)  ,
\end{equation}
where N is the number of beads per polymer, M the number of polymers, V the box volume, $m_k$ the mass of the k-th bead, $v_k$ the speed of the k-th bead, $F_{kl}$ the force between the k-th and the l-th bead and $r_{kl}$ their distance. $P_{\alpha \beta}$ is then averaged over a time $t_{avg}$~\cite{Lee2009a}. The autocorrelation was computed using the multiple-tau correlator method described in reference~\cite{Ramirez2010} and implemented in LAMMPS with the fix ave/correlate/long command. This method makes sure that the systematic error of the multiple-tau correlator to be always below the level of the statistical error of a typical simulation (see LAMMPS documentation). The viscosity $\eta$ of the system is then obtained by integrating $G(t)$ as,
$$ \eta = \int_{0}^{t \rightarrow \infty} G(t)dt $$

Given that our simulation are run for a finite time, the computed viscosity represents a lower bound for the true value. Such value is then obtained in simulation units $\frac{k_BT\tau_{Br}}{\sigma^3}$ where $\tau_{Br}=\frac{3\pi\eta_s\sigma^3}{k_BT} = 2.3 \mu s$.

To account for the noisy values appearing on large timestep values we model the behaviour of G(t) as a stretched exponential at large times. Specifically, we define $G(t) \approx a  e^{\left(-\frac{t}{\tau}\right)^b}$, and we fit a,$\tau$,b to approximate the exponential decay, starting at an arbitrarily found point, denoted as $t_e$. The viscosity is then obtained by numerical integration up to $t_e$, while the stretched exponential contribution is obtained by computing $\int_{t_e}^{\infty} a  e^{\left(-\frac{t}{\tau}\right)^b} = \frac{a\tau}{b}\Gamma\left(\frac{1}{b},(\frac{t_e}{\tau})^b\right)$, where $\Gamma(a,z)$ is the upper generalised gamma function. The sum between these two terms returns the viscosity estimate.

The Green-Kubo (GK) measurements are done in equilibrium in the fully-extruded cases. On the other hand, during transient extrusion polymers are out-of-equilibrium because LEFs are imposing non-thermal forces on them. Despite this, GK relations work in non-equilibrium steady-state systems in specific conditions~\cite{chun_nonequilibrium_2021}. To demonstrate that the loop extrusion \emph{per se} does not affect our GK calculation we selected four different timesteps separated by $2 \cdot 10^6 \tau_{Br}$ (significantly larger than the autocorrelation time of the squared radius of gyration in steady state, estimated to be approximately $5 \cdot 10^5 \tau_{Br}$) of a system with 10 LEFs per polymer and $\kappa_{off} = 10^{-8} \tau_{Br}$. We then blocked the extrusion process for each of these initial conditions, waited for $5 \cdot 10^6 \tau_{Br}$, sufficient to allow polymer equilibration and measured $G(t)$. We find that the GK stress-relaxation curve is independent on the specific state of the system once reached steady state, and is also identical to the one computed during the extrusion process (i.e. when it is not stopped). This means that the loop extrusion does not affect the viscoelasticity of the sample \emph{per se}, and that the most important contribution to the rheology is the change in polymer conformation. We argue that this is due to the extrusion regime we work in, as the extruded loops relax on timescales shorter than the time in between LEF steps~\cite{Chan2024pnas}. In turn, this leads to a virtually relaxed state in which polymer conformations are the main contribution to the viscoelastic properties. In other regimes, a correction to $G(t)$ would be needed to account for the energy-consuming non-equilibrium processes.

\subsection{Processivity}
We define a transient extrusion regime in which LEFs load/unload dynamically at rates $\kappa_{on}$ and $\kappa_{off}$, respectively. In this regime we define processivity as $\lambda = \frac{v_{LEF}}{k_{off}}$, but in many of our simulations the true speed of LEFs is hardly measurable. In fully extruded systems, since $k_{off}$ is zero, this value is assumed to be in average equal to the polymer length divided by the average number of LEFs per polymer. Instead, when $k_{off}>0$, we decided to compute the single LEF processivity and consider the average its distribution. Such distribution is found to be corresponding to a poissonian process, in agreement with our probability-dependent LEF removal.

\subsection{Modelling SMCs}
SMCs extrusion speed is of the order of $\qty{1}{\kilo\basepair\per\second}$~\cite{Ganji2018}. In our simulation the effective maximum speed is obtained by accepting all the moves with frequency $f_{eff}$, that for symmetric steps gives $v_{max} = 2 \sigma \cdot f_{eff} = 2\cdot10^{-3} \sigma \cdot \tau_{Br}^{-1} $ for $f_{eff} = 10^{-3} \tau_{Br}^{-1}$. Since each bead coarse grains $\sim \qty{30}{\basepair}$ of naked DNA, we have $v_{max} \sim \qty{26}{\kilo\basepair\per\second}$. However, this maximum speed is hardly reached during simulations because of the cutoff value. In system that allow transient binding of LEFs we can get an estimate of the effective extrusion speed by multiplying the average processivity by the unbinding rate $k_{off}$. For values of $\frac{\lambda}{\langle d \rangle } \sim 1$ this speed is found to be in the order of $10^{-5} \sigma \cdot \tau_{Br}^{-1}$.

\section{Results}

\subsection{Loop Extrusion drives compaction and dilution of entanglements}

\begin{figure*}[t!]
    \includegraphics[width=\textwidth]{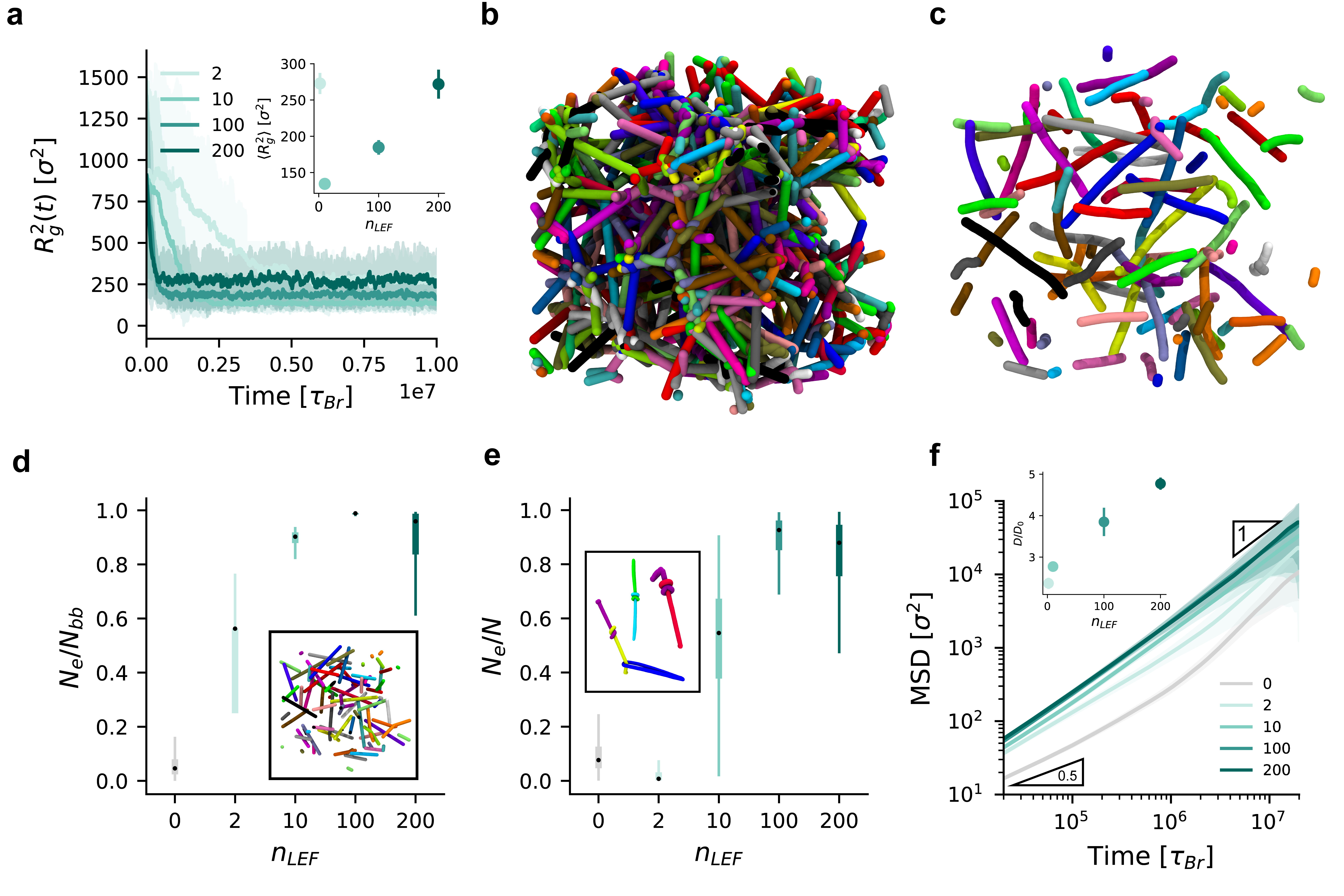}
    \caption{Equilibrium viscoelastic properties of monodisperse solutions with N=1000 for different values of $n_{LEF}$. \textbf{a} Average radius of gyration of polymers during extrusion and after equilibration. The faded area represents the width of the area covered by the distribution within the standard deviation. Inset shows the equilibrium value against $n_{LEF}$. \textbf{b-c} Snapshot of simulated systems after temperature quenching for a non extruded solution (\textbf{b}), extruded solution (\textbf{c}). \textbf{d} Box  plot representing the distribution of backbones entanglement lengths over the polymer length (number of extruders per polymer) obtained through PPA~\cite{Everaers2004} on a set of 5 uncorrelated replicas over 10 different time steps. The inset shows the final primite paths for $n_{LEF}=2$.
    \textbf{e} Box  plot representing the distribution of entanglement lengths over polymer length obtained through PPA~\cite{Everaers2004} on a set of 5 uncorrelated replicas over 10 different time steps. Inset displays examples of deadlocks found at $n_{LEF}=2$.
    \textbf{f} Average Mean Squared Displacement (MSD) of polymers' center of mass with its relative standard deviation. Inset displays the diffusion coefficient of the polymers' CoM normalised to the unextruded system as a function of $n_{LEF}$.
    }
    \label{fig:fig2}
\end{figure*}

\begin{figure*}[t!]
    \includegraphics[width=\textwidth]{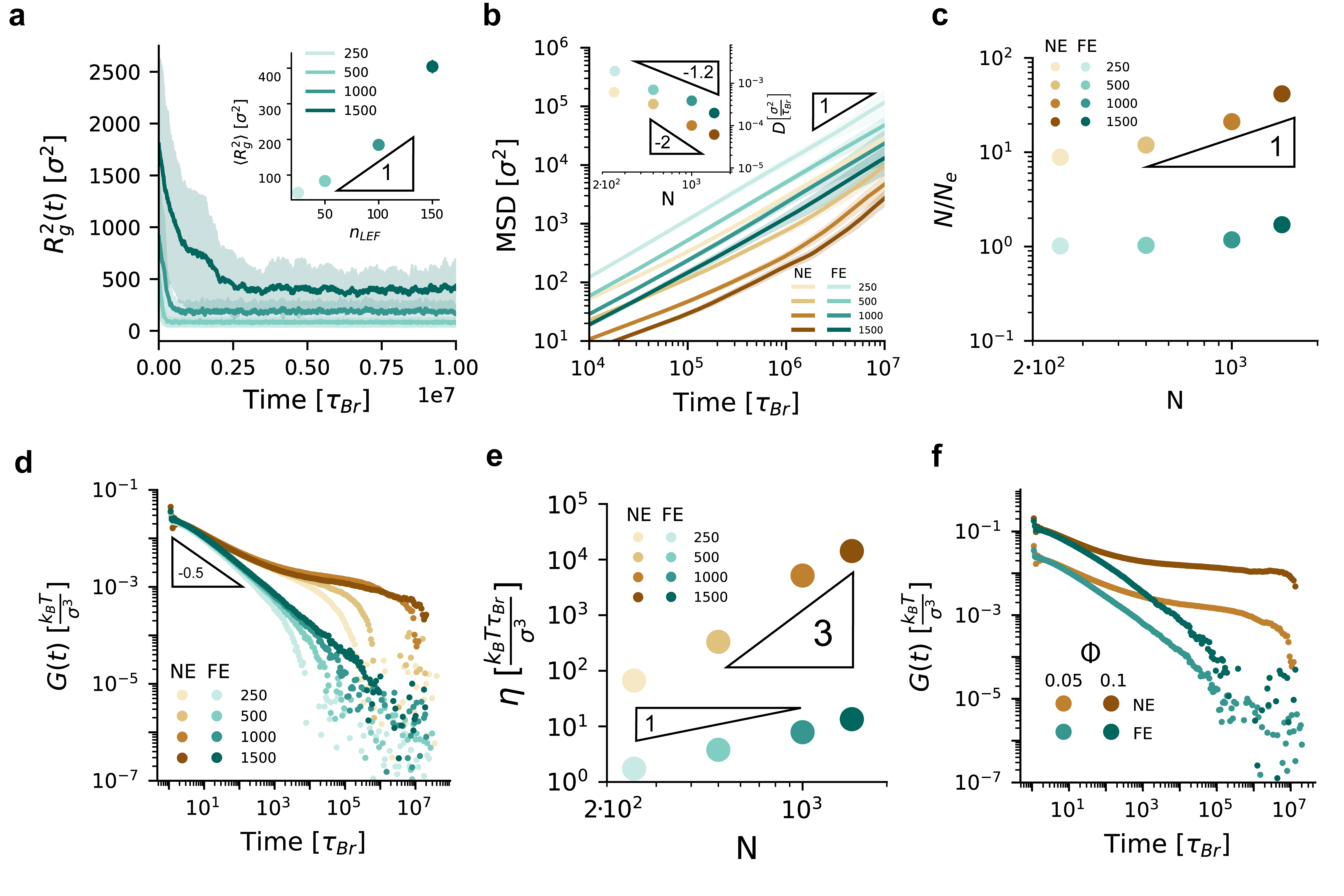}
    \caption{\textbf{a} Average radius of gyration of polymers during extrusion and after equilibration for three different values of N. The faded area represents the width of the area covered by the distribution within the standard deviation. Inset shows the equilibrium value against $n_{LEF}$. \textbf{b} Average Mean Squared Displacement (MSD) of polymers' center of mass with its relative standard deviation for non-extruded (NE) and fully extruded (FE) systems. Inset shows the diffusion coefficient of the polymers' centers of mass against the polymer length. \textbf{c} Number of entanglements for fully extruded (FE) and non-extruded (NE) configurations with different polymer length (from 250 to 1500 beads) and same LEF density ($n_{LEF}=100$). \textbf{d} Stress-relaxation function $G(t)$ of non-extruded (NE) and fully-extruded (FE) solutions in equilibrium. \textbf{e} Viscosity of extruded and non-extruded systems obtained by integration of the stress-relaxation function. \textbf{f} Stress-Correlation function of two 1000 beads long polymer systems with different volume fraction (0.05 and 0.1). Values in green represent fully extruded systems with $n_{LEF}=100$, while brown values correspond to non-extruded systems. } 
    \label{fig:fig3}
\end{figure*}

As explained in the Methods, we perform loop extrusion on dense and entangled polymer solutions. First, we qualitatively observe that the extrusion induces a geometrical deformation of the polymers into bottle-brush-like structures, as previously noticed~\cite{goloborodkoChromosomeCompactionActive2016,baniganLimitsChromosomeCompaction2019,polovnikov_crumpled_2023,Chan2024pnas} (Fig.~\ref{fig:fig1}c-f). Interestingly, this is accompanied by the onset of compartmentalisation and ``territories'', with the polymers becoming less intermingled~\cite{Racko2018} (see Fig.~\ref{fig:fig1}g-h). The radius of gyration $R_g^2$ displays a sharp reduction during the extrusion process (Fig.~\ref{fig:fig2}a). 
We then proceed to characterise the solution in equilibrium, when the squared radius of gyration plateaus and loop extrusion has stopped as LEFs are adjacent along the polymers. To make sure that our system is in equilibrium, we perform measurements once that the MSD has reached $\langle R_g^2 \rangle$ at steady state.
At large time, the steady state value $\langle R_g^2 \rangle$ displays a non-monotonic behaviour, which we understand as direct consequence of the bottle-brush structure (e.g., see snapshots in Fig.~\ref{fig:fig1}c-f): larger number of LEFs effectively induce a larger grafting density and longer backbones, which contribute to an increase of $R_g^2$. 
The observed stiffening is in agreement with both theory~\cite{markoPolymerModelsMeiotic1997} and experiments~\cite{Daniel2016} of synthetic bottle-brush polymers, as backbone stiffening is expected at larger grafting density due to increased entropic and stering interactions between the side chains. Additionally, from the snapshots of single chains in Fig.~\ref{fig:fig1} one can appreciate the change in polymer conformation as a function of the number of bound LEFs: from that of a dense comb polymer (small number of long side chains) to dense bottle-brush (large number of short side chains)~\cite{Daniel2016,Paturej2016}.

To understand how the conformational transition affects entanglements, we perform Primitive Path Analysis to the extruded solutions~\cite{Everaers2004,Uchida2008} (Fig.~\ref{fig:fig2}b-c). According to Ref.~\cite{Uchida2008} we can estimate the entanglement length $N_e$ as \begin{equation}
 N_e = l_K \left[ \left( c_\xi \rho_K l_K^3 \right)^{-2/5} + \left( c_\xi \rho_K l_K^3\right)^{-2} \right ] \end{equation}
 where $c_\xi = 0.06$, $l_K = 2l_p$ is the Kuhn length and $\rho_K = NM/(l_KL^3)$ is the number density of Kuhn segments. We find that non-extruded solutions have an average entanglement length $N_e = 96 \pm 65 \sigma$, somewhat larger than the theoretically expected value $N_e = 43 \sigma$ yet within error. 

Similarly to what was done in Ref.~\cite{Liang2019} for solutions of bottle-brush polymers, we first performed PPA on the polymer backbones by removing the side-loops (Fig.~\ref{fig:fig2}d). All the extruded solutions display significant disentanglement compared to the non-extruded control system ($n_{LEF}=0$). However, we expect side loops to be not entangled when the polymers display many short loops, so that they cannot thread each other~\cite{Xiong2023,Michieletto2016pnas}. To quantify the entanglement between side loops we performed PPA on the whole polymers. We note that in this case the overall chains are non-Gaussian. Still, the entanglement length displays an interesting behaviour: at small $n_{LEF}=2$ the system appears more entangled than the non extruded control case (Fig.~\ref{fig:fig2}e) and that $N_e$ grows larger only when more LEFs are bound to the polymers. By visually inspecting the simulations, we found that the reason behind this behaviour is the presence of deadlocks (see inset Fig.~\ref{fig:fig2}e). It is intriguing that deadlocks had been found previously in systems of ring polymers out-of-equilibrium~\cite{micheletti2024deadlock,oconnor2020}; it would thus be intriguing to understand if these deadlocks are also caused by the activity of the loop extruders. Moreover, both in Fig.~\ref{fig:fig2}d and e, solutions with $n_{LEF}=200$ appear to undergo a mild re-entanglement, arguably due to the backbone stiffening we previously observed.
These findings are in line with experiments on synthetic bottle-brush polymers~\cite{Daniel2016,Cai2015,Hu2011} which display greatly weakened entanglements with respect to their linear counterpart.

\subsection{Loop extrusion speeds up polymer dynamics}

Having quantified the dramatic change in the static properties of the solution, we expect to observe a similar large impact of loop extrusion on polymer dynamics. In Fig.~\ref{fig:fig2}f, we show that the mean squared displacement (MSD) of the polymers centre of mass (CoM) displays a significant speed up compared with the non-extruded case. More specifically, the diffusion coefficient of the polymers CoM, computed as $D = \lim_{ t \to \infty} \mathrm{MSD}/6t $ in the system with $N/n_{LEF} = 20$ is roughly 5 times larger than the control. 
Thus, while our results confirm the hypothesis that purely intrachain loop extrusion leads to an effective ``fluidification'' (or entanglement dilution) due to the transition from linear to bottlebrush-like structures~\cite{polovnikov_crumpled_2023,Chan2024pnas}, they also suggest that too many LEFs would drive re-entanglement in dense solutions, due to backbone stiffening. However, this re-entanglement does not affect the increase in diffusivity, a result that could be possibly due to a lower degree of interpenetration between side-chains and thus reduced friction~\cite{Abbasi2019}.

\subsection{Loop-extrusion-mediated fluidification is length-dependent}

We argue that this speed up in dynamics, or ``fluidification'', ought to be even more dramatic in more entangled systems, for instance in denser systems, or for longer chains. To test this, we performed simulations with shorter ($N=250$, $N=500$) and longer ($N=1500$) chains at fixed volume fraction ($\phi = 0.05$) and fixed LEF densities ($N/n_{LEF} = 10$), as reported in Fig.~\ref{fig:fig3}.
First, we discover that loop extrusion compacts all polymer lengths down to similar sizes and that the steady state $\langle R_g^2 \rangle$ scales roughly linearly with $n_{LEF}$ (Fig.~\ref{fig:fig3}a), which is consistent with the expected scaling~\cite{Paturej2016} $R_g^2 \sim L_{bb} L_{sl}^{1/2}$, where $L_{bb}$ is the length of the backbone (determined by the number of extruders) and $L_{sl}$ the length of side loops (determined by $N/n_{LEF}$). 

By comparing the dynamics of the CoM MSD, we observe that in all extruded solutions there is a clear loss of the early-time subdiffusive regime (Fig.~\ref{fig:fig3}b) as previously observed in solutions of bottlebrush polymers \cite{Abbasi2017}. For non-extruded configurations, the diffusion coefficient of the CoM scales close to the expected $D \sim N^{-2}$~\cite{Doi1988}. On the other hand, for extruded solutions we find $D \sim N^{-1.2}$.  Interestingly, at our optimal disentanglement density (1 LEF every 10 beads) we achieve total disentanglement for every tested configuration. This reflects on the average number of entanglements per polymer, displayed in Fig. \ref{fig:fig3}c. In entangled conditions this scales almost linearly with $N$, i.e. with the length of the linear chain, while after extrusion it is constant and close to 5, displaying similar level of entanglement independently on the polymer length.

To quantify the change in viscoelasticity due to the extrusion, we use the Green-Kubo relation and compute the autocorrelation of the off-diagonal components of the stress-tensor $G(t)$~\cite{Ramirez2010}. From these measure we estimate a value of entanglement plateau of $(1.6 \pm 0.1) \cdot 10^{-3} k_B T/\sigma^3$, from which, combining Eqs.~5-6 in Ref.~\cite{Uchida2008}, we obtain an effective entanglement length of $27 \pm 3 \sigma$, closer to the theoretical estimate $N_e = 43 \sigma$. The familiar entanglement plateau observed for linear polymers~\cite{Kremer1990} is completely lost in all extruded solutions, which instead follow a decay $G(t) \sim t^{-1/2}$, as for unentangled chains (Fig.~\ref{fig:fig3}d). We interpret this as a strong signature that most of the entanglements are lost, even for our longest polymers $N=1500$ which display $N/N_e \simeq 17$ in equilibrium. Perhaps more remarkably, the viscosity $\eta = \int_0^{\infty} G(t) dt$, typically scaling as $\eta \sim N^3$ for linear polymers, scales only linearly ($\eta \sim N$) in the fully extruded cases (Fig.~\ref{fig:fig3}e). The weak scaling of viscosity as a function of polymerisation index is in line with previous experimental results on densely grafted bottle-brushes~\cite{Hu2011}.

These findings strongly suggest that the fluidification mechanism is more dramatic in solutions of longer polymers as, $\eta_{LEF}/\eta_0 \sim N^{-2}$. For example, for our $N = 1500$ system, the viscosity of the extruded system is about 1000 times smaller than the non-extruded one.  Furthermore, we find even stronger fluidification in denser systems. To infer how the system density affects the extrusion-mediated viscosity reduction we proceeded to simulate a system of 50 polymers with $N = 1000$ and volume fraction $0.1$. The stress-correlation functions displayed in Fig.~\ref{fig:fig3}f displays identical properties between the two systems. However, the viscosity computed from such function shows a $\approx$2000-fold reduction in viscosity between the non-extruded and extruded configuration, showing a consistent increment from the $\approx$1000-fold decrease obtained for the less dense system. By extrapolating these results to genomic-size DNA, we expect the difference between the dynamics of chromosomes in presence/absence of (purely intrachain and non-bridging) LEFs will be several orders of magnitude. This implies that the large-scale rearrangement and dynamics of interphase and mitotic chromosomes are expected to be sensitive to the presence of active loop extrusion and may be tested in experiments via, e.g., displacement correlation  spectroscopy~\cite{Zidovska2013,gabrieleDynamicsCTCFCohesinmediated2022,Yesbolatova2022,Bruckner2023,Shaban2018a}.  

\begin{figure*}[t!]
    \includegraphics[width=\textwidth]{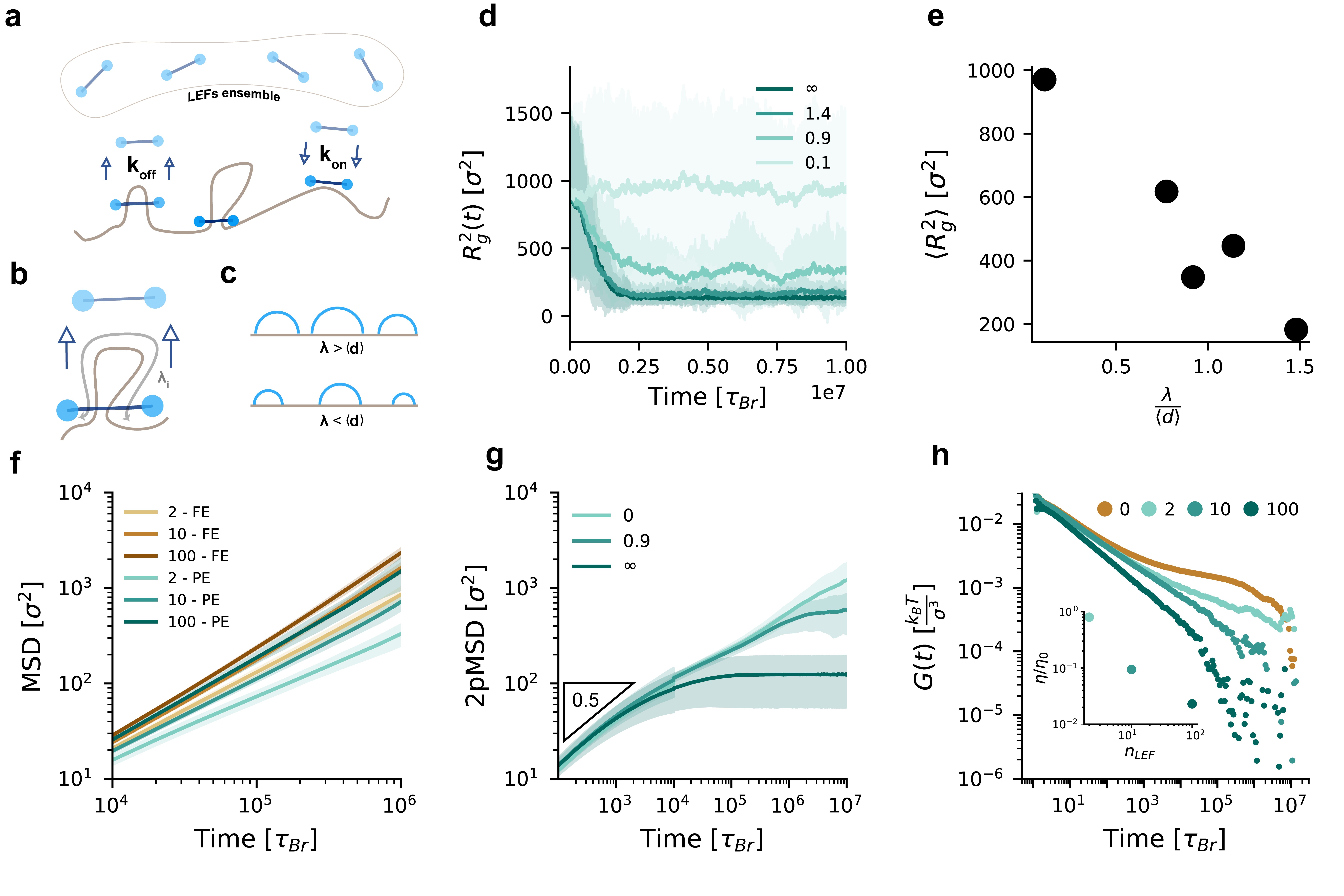}
    \caption{ \textbf{a} Sketch of our transient extrusion algorithm. LEF performs extrusion until they are removed with probability $k_{off}$, while non-binded LEFs can start extruding with probability $k_{on}$. \textbf{b} Sketch of how the extruded length for an SMC ($\lambda_i$) is estimated. \textbf{c} Sketch of extruded systems dependingly on the average processivity. \textbf{d} Average radius of gyration of polymers during extrusion and after equilibration for different values of $\lambda/\langle d \rangle$. In this context $\lambda/\langle d \rangle = \infty$ means that the polymer is fully extruded and $\lambda/\langle d \rangle = 0$ means that the polymer is unextruded.
    \textbf{e} Radius of gyration squared against normalised processivity. We multiply processivity by the number of LEFs per polymer and divide for the polymer length to get a collapse plot.
    \textbf{f} Average Mean Squared Displacement(MSD) for the center of mass of fully extruded (FE) and partially extruded polymers (PE) with  $\lambda/\langle d \rangle \simeq 1$ and different number of LEFs (2-100). \textbf{g} 2-point MSD taken as the average distance between two beads belonging to the same polymer with a contour distance of 600 beads with different values of $\lambda/\langle d \rangle$. \textbf{f} Stress-relaxation modulus of non-extruded and partially extruded systems with $\lambda/\langle d \rangle \simeq 1$ and different number of $n_{LEF}$. Inset shows the relative viscosity normalised to the viscosity of the non-extruded system. 
    }
    \label{fig:fig4}
\end{figure*}

\subsection{Out-of-equilibrium entangled solutions of transiently loop extruding polymers}

Having quantified the fluidification achieved in a fully-extruded steady state, we now turn our attention to the behaviour of our system under an out-of-equilibrium, active loop extrusion process for $N = 1000$. The total number of LEFs in the simulation is fixed, but the number bound at any one time fluctuate around the mean fraction $f_{bound} = \kappa_{on}/(\kappa_{on}+\kappa_{off})$. In the following, we focus on the so-called ``partially extruded'' case, in which the total extruded length is smaller than $N$, i.e. the LEFs kinetically unbind from the polymer before they can fully extrude the average distance between them (see Fig.~\ref{fig:fig4}a).
We highlight that the regime of partial extrusion is the one that appears to be most relevant in interphase~\cite{gabrieleDynamicsCTCFCohesinmediated2022,Michieletto2022tads}, and is characterised by the ratio between the average processivity and the average spacing between LEFs, i.e. $\lambda/\langle d \rangle = \langle v_{LEF} \rangle/\kappa_{off} \times N/n_{LEF}$ (see Fig.~\ref{fig:fig4}b). For instance, in partially extruded solutions $\lambda/\langle d \rangle < 1$.  

First, we find that the radius of gyration of the polymers depends on $\lambda / \langle d \rangle$: small values yield non-extruded polymers, while larger values yield a compaction similar to the fully extruded case (where $\lambda / \langle d \rangle \to \infty$ as $k_{off} = 0$). Interestingly, we also observe stable, intermediate compacted states in between a fully extruded and non-extruded state where the polymer is kept out-of-equilibrium by the kinetic binding of LEFs (Fig.~\ref{fig:fig4}d). 

In Fig. \ref{fig:fig4}e we display the distribution of the squared radius of gyration and connect its value to the average processivity of extruders. When the normalised processivity is very high ($>1$) we get systems with radius of gyration similar to the fully extruded ones. When the normalised processivity is small ($<1$) we instead get systems with radius of gyration close to the unextruded case. For intermediate values we find novel steady states with similar radius of gyration independently on the total number of SMCs in the system.

Despite being only partially extruded the average mean squared displacement for the center of mass of the polymers (Fig. ~\ref{fig:fig4}f) shows the loss of elastic behaviour at early times, while displaying also the LEF-dependent increase in diffusivity. However, the value of partially extruded-related MSDs is found to be always smaller than the correspondent fully-extruded MSDs.

To make a comparison with experiments, we compute the 2-point MSD~\cite{Yesbolatova2022,gabrieleDynamicsCTCFCohesinmediated2022}, i.e., the autocorrelation of the distance vector $\bm{d}$ between two given polymer segments separated by the curvilinear distance $l$ or $\textrm{2pMSD}(t,l) = \langle \left[ \bm{d}(t_0+t,s+l) - \bm{d}(t_0,s) \right]^2 \rangle_{t_0,s}$. In line with recent experiments~\cite{gabrieleDynamicsCTCFCohesinmediated2022,Bruckner2023,Yesbolatova2022} and Rouse theory, we observe a short-time scaling $\textrm{2pMSD}(t) \sim t^{1/2}$ that is surprisingly unaffected by loop the extrusion (Fig.~\ref{fig:fig4}g). 
A quantitative different behaviour of the $\textrm{2pMSD}(t)$ is seen only at large times due to the different polymer compaction, in agreement with experiments implementing rapid cohesin knockouts~\cite{gabrieleDynamicsCTCFCohesinmediated2022, Michieletto2022tads}. At intermediate timescales, both the non-extruded and partially extruded cases display $\textrm{2pMSD}(t) \sim t^{1/3}$, which is consistent with the one observed in experiments with embryo cells~\cite{Yesbolatova2022}. 

Finally, we compute the stress relaxation function $G(t)$ also for this out-of-equilibrium partially extruded scenario and once again we observe a clear lack of entanglement plateaus in the presence of LEFs (Fig.~\ref{fig:fig4}h). In turn, this translates in a significantly smaller effective viscosity of the partially extruded solution, which depends on the dimensionless ratio $\lambda/\langle d \rangle$. For partially extruded solutions ($\lambda/\langle d \rangle \simeq 1$) the viscosity is 100-fold smaller than the control, and we expect the same scaling with polymer length seen in Fig.~\ref{fig:fig3}. Importantly, we find that stopping the loop extrusion process and performing GK calculation yields the same $G(t)$ then the one obtained while loop extrusion is ongoing (not shown). This confirms that the loop extrusion \emph{per se} does not affect the viscoelasticity of the solution, but rather it does so indirectly through the change in polymer conformation. 

\section{Conclusions}

Motivated by the lack of understanding of how loop extrusion affects polymer entanglement and dynamics in dense solutions, we performed large-scale molecular dynamics simulations of entangled polymers under the action of loop extruding factors (LEFs). The first main discovery of our work is that loop extrusion, when it is exclusively intrachain, dramatically decreases the entanglement between chains in dense solutions. This result supports previous works suggesting that loop extrusion drives ``entanglement dilution''~\cite{polovnikov_crumpled_2023,Chan2024pnas}; we provide a solid numerical quantification of this effect through the use of PPA in dense solutions. Fully extruded chains display no sign of entanglement and, in fact, they self-organise into territories (Figs.~\ref{fig:fig1}-\ref{fig:fig2}). We then discovered that the mobility of the polymers is strongly sped up by the extrusion process, in turn yielding an effective fluidification of the solutions, leading to a much weaker scaling of viscosity $\eta \sim L$ (in contrast with $\eta \sim L^3$ for entangled systems) and a reduction up to 1000-fold for our longest polymers (Fig.~\ref{fig:fig3}). These effects are in line with the fact that synthetic bottle-brush polymers are more weakly entangled than their linear counterparts~\cite{Daniel2016}. Finally, we considered the case of partial extrusion using a transient binding model and showed that even in the partially extruded situation, with kinetic binding/unbinding of LEFs the viscosity of polymer solutions is greatly reduced via the active extrusion process (Fig.~\ref{fig:fig4}).   

Our findings suggest that loop extrusion may have a marked effect on the large-scale organisation and viscoelasticity of the nucleoplasm, through the change in conformation and dynamics of chromosomes. We note that our results are strongly dependent on the fact that we assumed loop extrusion to be exclusively intrachain  since we have neglected the formation of Z-Loops for computational simplicity. We argue that despite being observed in bacteria~\cite{Brando2021} and \textit{in vitro}~\cite{kim_dna-loop_2020}, there is no evidence of Z-loops formed in eukaryotes. We expect that considering interchain loop extrusion~\cite{bonatoThreedimensionalLoopExtrusion2021,Ryu2022nar} and bridging~\cite{Ryu2021} may drastically influence our results, eventually decreasing the mobility of the polymers and thus increasing the viscoelasticity of the solution. It will be interesting to test this scenario in the future.

\section*{Acknowledgements}
DM acknowledges the Royal Society and the European Research Council (grant agreement No 947918, TAP) for funding. The authors also acknowledge the contribution of the COST Action Eutopia, CA17139.  For the purpose of open access, the author has applied a Creative Commons Attribution (CC BY) licence to any Author Accepted Manuscript version arising from this submission.




\bibliography{bibliography}

\end{document}